\documentclass[12pt]{article}

\usepackage{newtxtext,newtxmath}

\usepackage{graphicx}
\usepackage{lineno}
\usepackage[utf8]{inputenc}
\usepackage[T1]{fontenc}
\usepackage{color}
\usepackage{ragged2e}
\usepackage{threeparttable}
\usepackage{booktabs}
\usepackage{booktabs, makecell, multirow, tabularx}
\usepackage{float}  
\usepackage{subfigure}  

\usepackage[letterpaper,margin=1in]{geometry}

\renewenvironment{abstract}
	{\quotation}
	{\endquotation}

\date{}

\makeatletter
\renewcommand{\fnum@figure}{\textbf{Figure \thefigure}}
\renewcommand{\fnum@table}{\textbf{Table \thetable}}
\makeatother

\usepackage{scicite}

\usepackage{url}

\def\scititle{
	High-Rate Free-Space Continuous-Variable QKD with Self-Referenced Passive State Preparation
}
\title{\bfseries \boldmath \scititle}

\author{
	Hanwen Yin$^{1\dagger}$,
	Xiaojuan Liao$^{1,\dagger}$,
    Yuehan Xu$^{1}$,
    Peng Huang$^{1,2,3,\ast}$,\and
    Kuntuo Zhu$^{4}$,
    Tao Wang$^{1,2,3}$,
    Guihua Zeng$^{1,2,3,5,\ast}$\and
        \small$^{1}$State Key Laboratory of Photonics and Communications \& Center for Quantum Sensing and Information Processing, Shanghai Jiao Tong University, Shanghai, 200240, China, \and
        \small Shanghai Jiao Tong University, Shanghai 200240, China \and
        \small$^{2}$Shanghai Research Center for Quantum Sciences, Shanghai 201315, China \and
        \small$^{3}$Hefei National Laboratory, Hefei 230088, China \and
        \small$^{4}$Shanghai Xiaoyuan Innovation Center, Shanghai, 201108, China \and
        \small$^{5}$Shanghai XunTai Quantech Co., Ltd, Shanghai, 200241, China \and
	\small$^\ast$Corresponding author. Email: huang.peng@sjtu.edu.cn, ghzeng@sjtu.edu.cn\and
	\small$^\dagger$These authors contributed equally to this work.
}

\begin{document} 
\maketitle

\begin{abstract} \bfseries \boldmath
Continuous-variable quantum key distribution (CVQKD) using passive state preparation (PSP) offers low-cost, high-rate secure communication. However, the existing PSP-CVQKD scheme with a transmitted local oscillator has high photon leakage noise and poor stability, making it unsuitable for high-loss transmission. In this work, for the first time, we propose and implement a local local oscillator PSP-CVQKD scheme, and give a theoretical proof of the equivalence of the PSP and GMCS protocol using temporal-mode theory. By employing the novel self-referenced pilot scheme to achieve high-precision time-varying frequency and phase compensation algorithms, we significantly improve the system’s signal-to-noise ratio and stability, and achieve an average secret key rate of 10.342 Mbps over a turbulent free-space channel with a maximum loss of 23.5 dB, which is even comparable with the classical active-state-preparation schemes. This innovative scheme is expected to advance the practical application of PSP-CVQKD technology and break new ground in low-cost and high-performance quantum communications. 
\end{abstract}

\section*{INTRODUCTION}
\noindent

Quantum key distribution (QKD) is principally classified into two categories: discrete-variable (DV) and continuous-variable (CV) realizations. Among these, CVQKD employs the orthogonal components of the coherent state for information encoding \cite{ref2,ref8,ref9}, thereby achieving a high key generation rate over relatively short transmission distances \cite{ref10,ref11,ref12} and demonstrating exceptional compatibility with classical optical communication systems \cite{ref13,ref14}. Specifically, it has been shown to suppress background noise with the assistance of the local oscillator (LO) within a coherent detection-based architecture \cite{ref15,ref16,ref17,ref18,ref19}. This property renders it well-suited for low-cost, chip-integrated photonic applications and has emerged as a prominent area of interest in quantum communication research \cite{ref13}.

The Gaussian modulated coherent state (GMCS) CVQKD has been proven to be reliable and secure against general attacks \cite{ref20,ref21,ref22}, and it has been extensively validated in both laboratory and real outside environments \cite{ref23,ref24,ref25,ref26,ref27,ref28,ref29}. And the local LO (LLO) architectures are particularly important in GMCS-CVQKD implementations because they eliminate the need to transmit LO with quantum signals, improving system stability, and avoiding multiple practical security attacks \cite{secure_p_1,secure_p_2}. However, the GMCS protocol still faces the challenges of high-rate modulation and high stability in practical applications. Despite the employment of high-speed systems with modulation rates of up to 10 Ghz in numerous contemporary studies \cite{ref28,ref29,Hajomer_10G}, the integration of high-speed active modulation devices, such as the arbitrary waveform generator, remains a technical constraint for the practical implementation of QKD. This technical limitation directly restricts the improvement of the secret key rate (SKR) under the active modulation architecture.

Passive-state-preparation (PSP) scheme \cite{ref30,ref31,ref32,ref33} is a method of CVQKD that utilizes the internal field fluctuation properties of the thermal source, combined with beam splitters, optical attenuators, and detectors, to prepare a quantum state and distribute secret keys. The PSP-CVQKD scheme differs from the conventional ASP scheme with quantum random number generator (QRNG) and electro-optic modulator by utilizing the intrinsic quadrature fluctuations of a single mode thermal state to generate quantum random numbers \cite{ref30}. This innovative scheme has the potential to significantly reduce system costs and simplify hardware architecture, making it a promising solution for practical applications in fields such as free-space and fiber communications.

Although some important progress has been made in PSP-CVQKD research, such as the experimental validation in optical fiber and atmospheric channels by transmitted LO (TLO) \cite{ref34,ref35}, there are still some urgent improvements in the existing schemes. Among them, although the TLO scheme achieves functional verification to a certain extent, it brings problems such as increased photon leakage noise, decreased system stability, and increased security risk, and the TLO scheme is not very suitable for long-distance transmission scenarios. These problems affect the efficiency and security of key distribution and become the key obstacles limiting the practical application of PSP-CVQKD. To cope with these problems, a theoretical scheme of LLO-PSP-CVQKD multiplexing has been proposed, but not experimentally verified \cite{ref44}.

In this paper, we propose and successfully realize a new LLO-PSP-CVQKD scheme. Quite different from the traditional TLO scheme, our scheme sets the LO at the receiver, which fundamentally avoids the mutual interference between the signal light and the LO light in the transmission process. By designing a classical light-assisted optical structure, we fundamentally overcome the problem of thermal signals occupying the entire frequency band without being able to realize LLO, and significantly improve the signal-to-noise ratio (SNR) and stability of the system. In terms of technical realization, we have developed self-referenced time-varying frequency and phase noise compensation algorithms for LLO-PSP-CVQKD, which further improve the overall performance of the system, and achieved time-varying channel transmittance estimation through a simpler optical structure. The proposed LLO-PSP-CVQKD scheme demonstrates remarkable performance in long-distance communication scenarios, achieving an average asymptotic key rate of 10.342 Mbps even under challenging turbulent channel conditions with transmittance fluctuations ranging from -16 dB to -23.5 dB. This outstanding result highlights the significant advantage of the LLO scheme in maintaining high-speed secure key distribution over extended distances where channel transmittance typically suffers severe degradation.

\section*{RESULTS}

\subsection*{Continuous-time Mode Theory for LLO-PSP-CVQKD Scheme}

\begin{figure*}[ht]
\centering
\includegraphics[width=1\linewidth]{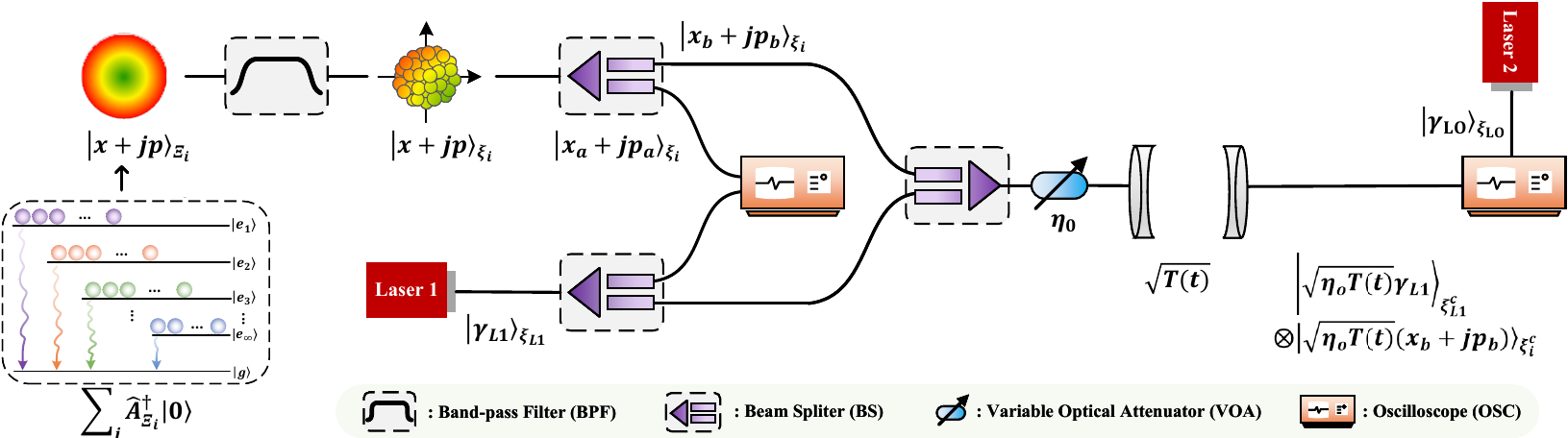}
\caption{Schematic of LLO-PSP-CVQKD theory. In a thermal source, photons at different energy levels produce light in different modes through spontaneous radiation. After continuumization, this light can be equated to GMCS. The signals in the available bandwidth range are selected by band-pass filtering to obtain coherent states that can carry information. The single-mode laser and the thermal light are separately beam split, and a part of the split beam is detected at Alice, and the other part is transmitted to Bob. In the receiver, the laser generates the LO light for detecting the received signal.}
\label{fig:PSPmodel}
\end{figure*}

In discrete-mode quantum optics, thermal states  are considered Fock states of classical probability superposition, its density matrix can be represented as
\begin{equation}
\begin{aligned}
		\hat{\rho}=\sum_n P(n) \left| n \right \rangle \left \langle n \right |,
\end{aligned}
\end{equation}
where $P(n)=\overline{n}/\left(1+\overline{n}\right)^{n+1}$ follows the Bose-Einstein distribution. According to the Weisskopf-Wigner theory of the spontaneous emission \cite{ref37,ref36} (more details in Supplementary Information, Note 1), the continuum of the light field can be provided by the following process
\begin{equation}
	\begin{aligned}
		\left| \gamma \right \rangle=\int_{\lambda} d \lambda \frac{g^*_{\lambda} }{\delta_{\omega_{\lambda}}+j \Gamma/2} \left| 1 \right \rangle_{\lambda},
	\end{aligned}
\end{equation}

Based on continuous-mode quantum optics \cite{ref38,ref39,ref40,ref41,ref42,ref43}, we rewrite the above results in the following form
\begin{equation}
	\begin{aligned}
		\left| \{\gamma\} \right \rangle=\sum_i \hat{A}^{\dagger}_{\Xi_i} \left| 0 \right \rangle,
	\end{aligned}
\end{equation}
where $\Xi_i \left(t\right)=\Xi^{0}_{i}\left(t\right) e^{-j \omega_i t}$ is the temporal-mode field operator, and the weight is $\Xi^{0}_{i}\left(t\right)=\frac{g^*_i}{\delta_{\omega_i}+j \Gamma/2}$. After the continuum, the density matrix of the thermal state can be represented as
\begin{equation}
\begin{aligned}
		\hat{\rho}=\otimes_i \sum_n P(n) \left| n \right \rangle_{\Xi_i} \left \langle n \right |_{\Xi_i}.
\end{aligned}
\end{equation}
Therefore, a single-mode thermal state can be equivalent to a coherent state with a complex amplitude, whose randomness is provided by classical probability
\begin{equation}
\begin{aligned}
		\left|x+jp\right \rangle_{\Xi_i}&=e^{-V_{\rm{A}}/4} \sum_n \frac{\left(x+jp\right)^n}{n!} \left(\hat{A}^{\dagger}_{\Xi_i}\right)^n \left| 0 \right \rangle \\
		&=\sum_n\sqrt{P(n)} \left| n \right \rangle_{\Xi_i},
\end{aligned}
\end{equation}
where $x$ and $p$ form a complex amplitude that obeys the Bose-Einstein distribution of $P(n)$, $V_{\rm{A}}=2|x+jp|^2$ is the variance. When the number of photons is large ($n \gg 1$), the distribution can be approximated as a Poisson distribution $P(n)=e^{-\overline{n}}\overline{n}^n/n!$. At this time, the quantum state can be effectively regarded as a GMCS. Although we have obtained the GMCS here, since the spectral range of each mode is not specified, in fact, this GMCS cannot carry information.

Figure~\ref{fig:PSPmodel} presents the schematic of the continuous-time mode theory of the LLO-PSP-CVQKD. Depending on the spectral range used, perform re-orthogonalization and re-normalization. A simple operation is to perform a bandpass filtering (BPF) operation on the thermal state. Taking the spectral center $\omega_s$ as the reference, the field operator can be rewritten as
\begin{equation}
\begin{aligned}
		\xi_i \left(t\right)=&\xi^{0}_{i}\left(t\right) e^{-j \omega_s t},\\
        \xi_i^0(t)=&\frac{g_i^*}{\delta_{\omega_i}+j\Gamma/2},
\end{aligned}
\end{equation}
and the equivalent coherent state is $\left|x+jp\right\rangle_{\xi_i}$. The quadratures $x$ and $p$ as the time-varying parameter has a repetition rate (modulation rate in GMCS) $F_m = 2(\omega_e - \omega_s)$, where $\omega_e$ is the maximum margin of thermal state.

A feature of the PSP scheme is to replace the coherent state preparation with the detection of the coherent state, i.e., the thermal light generated by the amplified spontaneous emission (ASE) source is divided by a beam splitter (BS) with transmittance $T_1$, the process is given by
\begin{equation}
    \begin{aligned}
        &\hat{B}_{\xi_i}\left[cos^{-1}\sqrt{T_1}\right]\cdot \left|x+jp\right\rangle_{\xi_i} \\
    =& \left|\sqrt{T_1}\left(x+jp\right)\right\rangle_{\xi_i}\otimes \left|\sqrt{1-T_1}\left(x+jp\right)\right\rangle_{\xi_i}\\
    =&\left|x_a+jp_a\right\rangle_{\xi_i}\otimes\left|x_b+jp_b\right\rangle_{\xi_i},
    \end{aligned}
\end{equation}
$\left|x_a+jp_a\right\rangle_{\xi_i}$ is detected at Alice. $\left|x_b+jp_b\right\rangle_{\xi_i}$ enters the subsequent optical path to be further transmitted into the channel with the beacon light.

At Alice’s side, the photon-wavepacket coherent state generated by Laser 1 is $\left|\gamma_{\mathrm{L1}}\right\rangle_{\xi_{\mathrm{L1}}}$, where \(\xi_{\mathrm{L1}}(t)=\xi_{\mathrm{L1}}^0(t)e^{-\mathrm{j}\omega_{s}t}\), $\omega_s$ is the optical carry frequency of Laser 1. A portion of the Laser 1 light enters Alice’s integrated coherent receiver (ICR) as LO for coherent detection. When the detector and signal bandwidths are the same, a single-point sample can represent the final data \cite{ref42}. Then, in the receiver model with an heterodyne detector followed by a filter, the impulse response function of the filter approximates the wavepacket of the measured state. We can rewrite the measurement by a Gram-Schmidt process:
\begin{equation}
    \label{eq2.8} \hat{I}_{\Xi_{\mathrm{dsp}}^A}^{\mathrm{SNU}}=\hat{X}_{\Xi_{\mathrm{dsp}}^A}+\mathrm{j}\cdot \hat{P}_{\Xi_{\mathrm{dsp}}^A}=\sqrt{\eta_e\eta_A}\left(\hat{X}_{\xi_i}+\mathrm{j}\cdot\hat{P}_{\xi_i} \right),
\end{equation}
where $\eta_e$ is the quantum detection efficiency of ICR, and $\mathrm{SNU}$ is the shot noise unit. Alice only detects the signal without digital signal processing (DSP) post-processing operation, so $f_{\mathrm{dsp}}^A=\frac{{\xi_i^{0^*}}(t)}{\xi_{\mathrm{L1}}^{0^*}(t)}$. The mode-matching factor $\sqrt{\eta_A}$ is:
\begin{equation}
    \label{eq2.9}
    \begin{aligned}
        \sqrt{\eta_A}=&\int_{t_0}^{t_0+N/F_m}dt\frac{1}{\sigma_{\mathrm{cal}}^*}\xi_{\mathrm{L1}}^*(t)G_{\mathrm{dsp}}^{N,A^*}\xi_i(t)\\
        =&\int_{t_0}^{t_0+N/F_m}dt\frac{1}{\sigma_{\mathrm{cal}}^*}\xi_{\mathrm{L1}}^{0^*}(t)\xi_i^0(t)\frac{{\xi_i^{0^*}}(t)}{\xi_{\mathrm{L1}}^{0^*}(t)}=1.
    \end{aligned}
\end{equation}
The detected signal $\mu_A$ is
\begin{equation}
    \label{eq2.10}
    \begin{split}
        \mu_A
        =&\left\langle x_a+jp_a\right|\hat{I}_{\Xi_{\mathrm{dsp}}^A}^{\mathrm{SNU}}\left| x_a+jp_a\right\rangle_{\xi_i}\\
        =& \sqrt{\eta_e}\left(x_a+jp_a\right).
    \end{split}
\end{equation}
Laser 1 splits out part of the light as beacon light to merge with $\left|x_b+jp_b\right\rangle_{\xi_i}$, and the beacon light calibrates the center frequency of the thermal light. The combined light enters the variable optical attenuator (VOA) with an attenuation coefficient of $\eta_0$ for power adjustment, and the final signal sent to Bob is
\begin{equation}
    \left|\sqrt{\eta_0}\gamma_{\mathrm{L1}}\right\rangle_{\xi_{\mathrm{L1}}}\otimes\left|\sqrt{\eta_0}\left(x_b+jp_b\right)\right\rangle_{\xi_i}.
\end{equation}

Subsequently, the signal enters a free-space channel for transmission. During this process, it will experience phase rotation, power attenuation, and polarization change. The time-varying phase rotation caused by the free-space channel can be expressed as \(\theta_c^t\). Considering the linear, time-varying transmittance of the free-space channel, the transmittance can be expressed as \(T(t)\). The phase rotation acting on the envelope and the transmittance decay acting on the number of photons are considered here. Thus, the photon-wavepacket coherent state reaching Bob can be expressed as
\begin{equation}
\label{eq2.12}
    \left| \sqrt{\eta_0T(t)}\gamma_\mathrm{L1}\right\rangle_{\xi_{\mathrm{L1}}^c}\otimes\left| \sqrt{\eta_0T(t)}\left(x_b+jp_b\right)\right\rangle_{\xi_i^c}.
\end{equation}
The photon-wavepacket, after passing through the channel, is
\begin{equation}
\label{eq2.13}
    \begin{aligned}
        \xi_i^c(t) & = \xi_i^0(t)e^{\mathrm{j}\left(-\omega_st+\theta_c^t\right)},\\
        \xi_\mathrm{L1}^c(t)&=\xi_\mathrm{L1}^0(t)e^{\mathrm{j}\left(-\omega_st+\theta_c^t\right)}.
    \end{aligned}
\end{equation}

Upon reaching Bob, the photon-wavepacket coherent state first passes through the polarization controller (PC) for polarization correction. Assuming that the polarization controller is ideal, polarization leakage is not considered. Bob's Laser 2 produces another photon-wavepacket coherent state denoted as $\left|\gamma_{\mathrm{LO}}\right\rangle_{\xi_{\mathrm{LO}}}$. Here, the LO produced by Laser 2 is represented by the expression $\xi_{\mathrm{LO}}(t)=\xi_{\mathrm{LO}}^0(t)e^{\mathrm{j}\left(-\omega_{\mathrm{LO}}^tt+\theta_{\mathrm{LO}}^t\right)}$. The variables $\omega_{\rm{LO}}^t$ and $\theta_{\rm{LO}}^t$ represent the optical carrier frequency of the LO and the time-varying phase difference between Laser 1 and Laser 2, respectively. Utilizing heterodyne detection, Bob inputs these states into the ICR with a detection efficiency of $\sqrt{\eta_e}$ for the two orthogonal measurement modes. Bob's ICR detection results can be expressed as
\begin{equation}
    \label{eq2.14}
    \begin{aligned}                 \hat{I}_{\Xi_{\mathrm{dsp}}^B}^{\mathrm{SNU}}&=\hat{X}_{\Xi_{\mathrm{dsp}}^B}+\mathrm{j}\cdot \hat{P}_{\Xi_{\mathrm{dsp}}^B}=\sqrt{\eta_e\eta_B}\left(\hat{X}_{\xi_i^c}+\mathrm{j}\cdot\hat{P}_{\xi_i^c} \right),\\
    \hat{I}_{\Xi_{\mathrm{dsp}}^q}^{\mathrm{SNU}}&=\hat{X}_{\Xi_{\mathrm{dsp}}^q}+\mathrm{j}\cdot \hat{P}_{\Xi_{\mathrm{dsp}}^q}=\sqrt{\eta_e\eta_q}\left(\hat{X}_{\xi_\mathrm{L1}^c}+\mathrm{j}\cdot\hat{P}_{\xi_\mathrm{L1}^c} \right).
    \end{aligned}
\end{equation}

\subsection*{Protocol and setup}

\begin{figure*}[ht]
\centering
\includegraphics[width=0.8\linewidth]{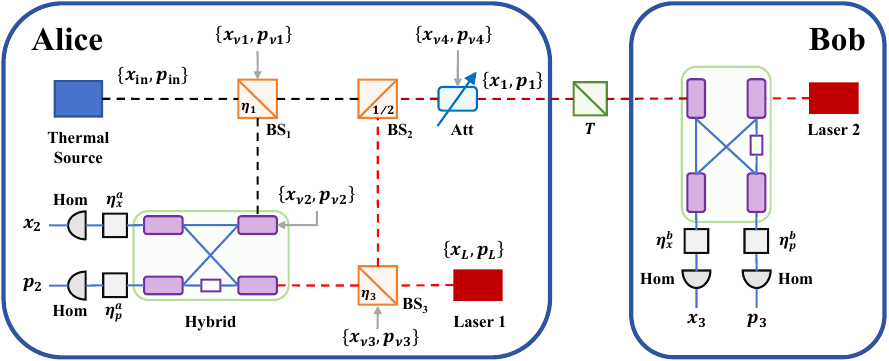}
\caption{Experimental schematic of LLO-PSP-CVQKD. $\mathrm{BS}_1$ and $\mathrm{BS}_3$: 99:1 beam splitter. $\mathrm{BS}_2$: 50:50 beam splitter. Att: optical attenuator with attenuation factor $\eta_0$. T: free-space channel transmittance. Hom: homodyne detector. $\{x_{\nu},p_{\nu}\}$: vacuum noise. $\eta_x$ and $\eta_p$: detection efficiency.}
\label{fig:sche}
\end{figure*}

The schematic of the experimental LLO-PSP-CVQKD scheme is presented in Figure~\ref{fig:sche}. A distinctive feature of the PSP scheme is the replacement of coherent state preparation with coherent state detection. Specifically, the thermal source output ($x_{in}$ and $p_{in}$) at Alice is separated into two spatial modes. One of the modes is then directly detected by heterodyne detection at Alice to obtain the quantum random numbers ($x_2$ and $p_2$), which are generated by the intrinsic fluctuations of the thermal source. The other mode is then combined with the beacon light, attenuated by optical attenuator $\mathrm{Att}$, and transmitted to Bob via a free-space channel. In this configuration, the beacon light is generated by a single-mode Laser 1, and the center frequency of Laser 1 is equal to the frequency of the thermal signal, thereby calibrating the frequency of the thermal signal. Subsequently, Bob uses another LO from Laser 2 to measure the input quantum states transmitted from Alice with heterodyne detection and obtain the received signals ($x_3$ and $p_3$). Following the completion of the frequency offset and phase compensation, frame synchronization, and channel transmittance estimation steps, Alice and Bob proceed to parameter estimation. Finally, the security key can be generated by reconciliation and privacy amplification operations. Since Alice obtains information about the quadrature components of the sent state using heterodyne detection, this process inevitably introduces excess noise due to PSP. The derivation of the PSP noise introduced in this experiment is described in detail in the Methods section.

\begin{figure*}[ht]
\centering
\includegraphics[width=1\linewidth]{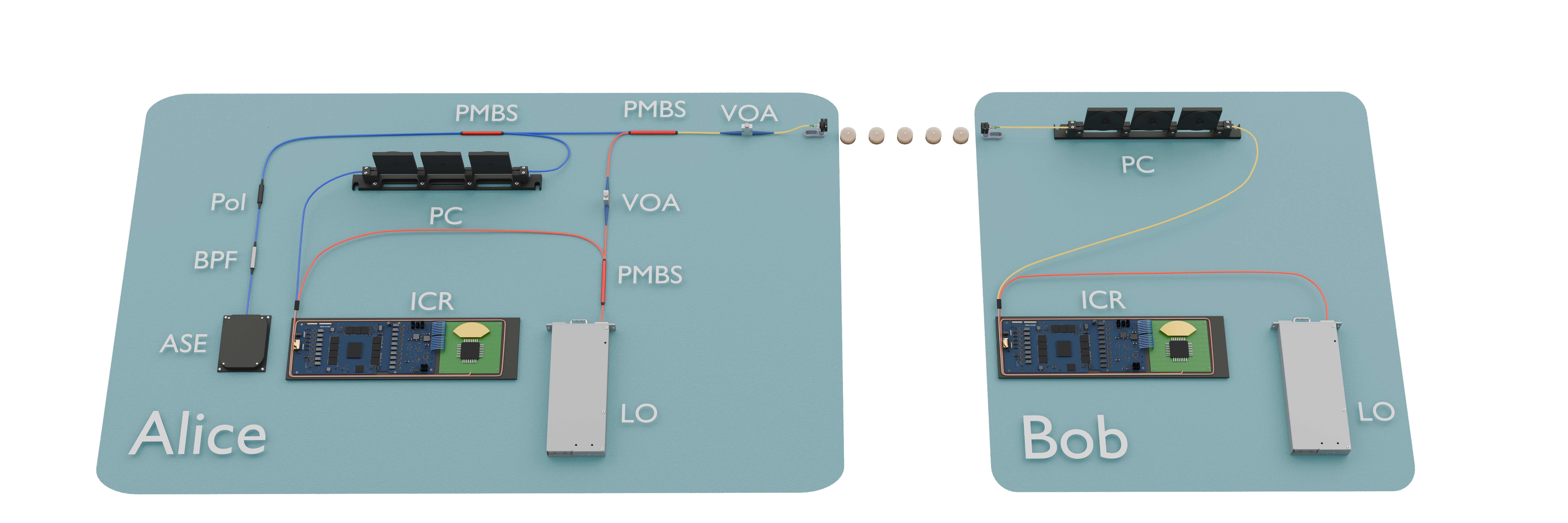}
\caption{LLO-PSP-CVQKD optical layout. ASE, amplified spontaneous emission; BPF, bandpass filter; Pol, polarizer; PMBS, polarization-maintaining beam splitter; PC, polarization controller; ICR, integrated coherent receiver; LO, local oscillator; VOA, variable optical attenuator.}
\label{fig:setup}
\end{figure*}

\begin{figure*}[ht]
    \centering
    \includegraphics[width=1\linewidth]{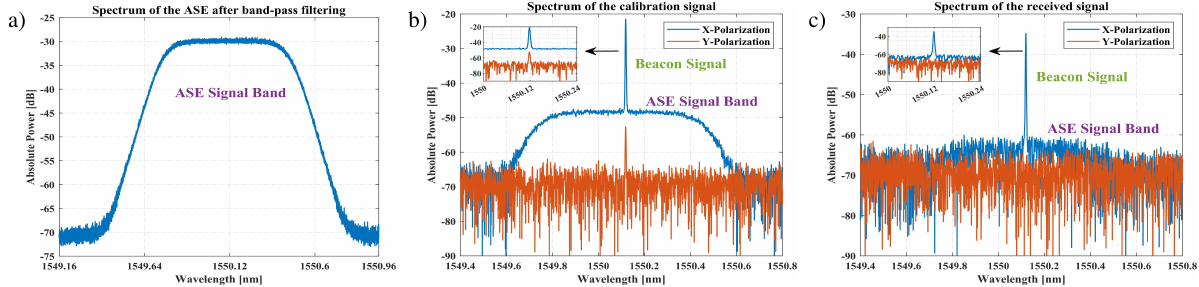}

    \caption{Spectra of the system at different nodes. (a) Spectra of the ASE source after bandpass filtering. The center is 1550.12 nm, and the bandwidth is 0.4 nm. (b) Spectrum of the signal before entering the channel. Maximum power in the X-polarization direction means that the polarization of the transmitted signal is optimal. (c) The spectrum of the signal entering the ICR detector. The signal is adjusted to the polarization optimum by the polarization controller.}
    \label{fig:spectrum}
\end{figure*}

The experimental configuration of the LLO-PSP-CVQKD system is illustrated in Figure~\ref{fig:setup}. In this system, a commercial ASE light source is utilized to generate a broadband thermal source. The spectrum of the thermal source is a multimode broadband spectrum. The thermal-state signal output from the ASE source first passes through an optical BPF with a central wavelength of 1550.12 nm and a bandwidth of 0.4 nm. This operation is intended to eliminate the modes that do not correspond to the quantum signal. Meanwhile, this narrow-bandwidth BPF also has the function of suppressing multi-mode noise. The spectrum of the thermal signal after BPF is illustrated in Figure~\ref{fig:spectrum}(a). The central frequency of the thermal-state signal is 1550.12 nm, and the spectral width is 0.4 nm. A second-order correlation function is then performed on the filtered thermal source, and the results of 1000 tests are shown in Supplementary Information, Note 2. The measured value of the normalized second-order correlation function fluctuates slightly around 2. This result is consistent with the theoretical expectation of the thermal source, indicating that the thermal source can be considered approximately an ideal perfect thermal source at this time (more details in Supplementary Information, Note 3).

The thermal signal, after BPF, is shaped into linearly polarized light by a polarizer (Pol) and input into a 99:1 polarization-maintaining BS (PMBS). 99\% of the signal passes through a PC to properly adjust the polarization and into an ICR with a 20 GHz detection bandwidth on the Alice. The residual 1\% of the signal is directed to the PMBS, where it is combined with the beacon light. The light from Laser 1 with a linewidth of <100 Hz at the Alice has a central wavelength of 1550.12 nm, matching the central wavelength of the ASE signal. A 90:10 BS passes the single-mode laser emission from Laser 1, with 90\% serving as the LO and entering the ICR to detect random numbers at Alice, and the remaining 10\% serving as the beacon light. A VOA adjusts the intensity to an appropriate level, and this is combined with the ASE signal. A VOA then adjusts the signal appropriately for equivalent modulation variance before being sent into the free-space channel. Figure~\ref{fig:spectrum}(b) shows the spectrum of the signal sent to Bob. The central frequency of the beacon light is 1550.12 nm, which accurately calibrates the central frequency of the thermal-state signal. In additinn, we introduced a burning candle in the free-space channel to simulate atmospheric turbulence.

After receiving the signal, Bob first adjusts the polarization state of the signal appropriately using a PC and then inputs the signal into an ICR with a 20 GHz detection bandwidth for coherent detection. The LO at Bob's side is generated by Laser 2 with a central wavelength of 1550.13 nm, which is aimed at reducing the 1/f noise contained in the beat signal. Figure~\ref{fig:spectrum}(c) shows the spectrum of the signal received by Bob. It is noteworthy that the employment of the 20 GHz ICR detector has facilitated the realization of a high-speed modulation equivalent to 20 GHz, a level that significantly surpasses that of the ASP CVQKD.

\subsection*{Excess noise suppression}

Maintaining lower levels of excess noise is a necessary prerequisite for achieving high performance. In the PSP-CVQKD scheme, the PSP noise constitutes a unique excess noise component that distinguishes it from the ASP-CVQKD scheme. In contrast to the  TLO-PSP-CVQKD scheme, the LLO scheme does not need to consider the photon leakage noise due to the strong LO light, but it does need to take into account the noise introduced by the frequency shift between the signal light and the localization light. Under the LLO-PSP-CVQKD scheme, a complete noise model can be constructed, which is capable of describing the system noise scenario as follows:
\begin{equation}
    \label{eq3.1}
    \varepsilon = \varepsilon_{\mathrm{psp}}+\varepsilon_{\mathrm{freq}}+\varepsilon_{\mathrm{phase}}+\varepsilon_{\mathrm{chan}}+\varepsilon_{\mathrm{fad}}+O\left(\varepsilon_{\mathrm{mode}}\right),
\end{equation}

\noindent where $\varepsilon_{\mathrm{psp}}$, $\varepsilon_{\mathrm{freq}}$, $\varepsilon_{\mathrm{phase}}$, $\varepsilon_{\mathrm{chan}}$, $\varepsilon_{\mathrm{fad}}$ and $\varepsilon_{\mathrm{mode}}$ characterize the excess noise induced by passive state preparation, frequency offset, phase drift, fluctuation of transmittance, channel instability and out-of-band optical mismatch, respectively. Among them, the excess noise introduced by out-of-band optical mismatch $O\left(\varepsilon_{\mathrm{mode}}\right)$ is extremely small and can be ignored in practical analysis and calculation. $\varepsilon_{\mathrm{channel}}$ is closely related to environmental factors, and it is difficult to effectively control it under the existing technical conditions. Therefore, the focus of the excess noise suppression is to suppress $\varepsilon_{\mathrm{psp}}$, $\varepsilon_{\mathrm{freq}}$, $\varepsilon_{\mathrm{phase}}$, $\varepsilon_{\mathrm{fad}}$. $\varepsilon_{\mathrm{psp}}$ caused by the LLO-PSP-CVQKD scheme can be calculated by the following equation (see Methods for detailed derivation):
\begin{equation}
    \label{eq3.2}
    \begin{split}
        \varepsilon_{\mathrm{psp}} = 
        &\frac{-\eta_x^a\left[ 2(V_A+\eta_0)\sqrt{\eta_1(1-\eta_1)}-\eta_0\sqrt{\eta_3(1-\eta_3)}\right]^2}{2V_A\eta_x^a(1-\eta_1)+\eta_0\eta_x^a(1-\eta_3)+2\eta_0(\nu^{\mathrm{ele}}_a+1)}\\
        &+2V_A\eta_1+\eta_0\eta_3,\\
    \end{split}
\end{equation}
\noindent where $V_A=\eta_0n_0$ is the equivalent modulation variance, which can be obtained by pre-calibration before the experiment, $n_0$ is the average number of photons output from the thermal source, and the magnitude of $V_A$ can be changed by adjusting the attenuation factor $\eta_0$ of the attenuator, $\eta_1$ and $\eta_3$ are the beam splitter ratios, $\eta_x^a$ is the quantum efficiency of the detector, $\nu^{\mathrm{ele}}_a$ is electrical noise of Alice's detector. Accordingly, the beam splitting ratio $\eta_1$ and attenuation factor $\eta_0$ can be optimized to reduce the $\varepsilon_{\mathrm{psp}}$.

\begin{figure*}[ht]
    \centering
    \includegraphics[width=\linewidth]{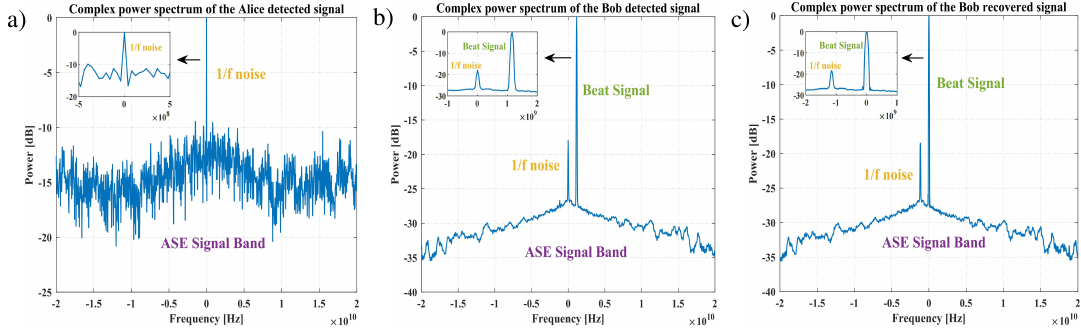}
    \caption{Complex power spectra (CPS) of the system at different nodes. (a) CPS of the Alice detected signal. (b) CPS of the Bob detected signal. (c) CPS of the Bob recovered signal.}
    \label{fig:3-3}
\end{figure*}

The complex power spectrum (CPS) of the ASE signal detected by Alice is presented in Figure~\ref{fig:3-3}(a). Limited by the detection bandwidth of the ICR detector, the bandwidth of the thermal signal is located at 0 Hz, and the spectral width is 20 GHz. Figure~\ref{fig:3-3}(b) shows the CPS of the signal detected by Bob, and the beacon light, after mixing with Bob's LO light, is presented as a beat signal within the detection bandwidth, and this beat signal reflects the center of the ASE signal. In the ideal case, if the center frequency of Bob's LO light is perfectly aligned with the center frequency of the ASE signal, then the beat signal should be at 0 Hz in the spectrum, which corresponds to the center of the received ASE signal at 0 Hz. However, at this time, the beat signal appears at about 1.1 GHz, which means that the ASE signal has a frequency deviation of 1.1 GHz compared with that of the signal at Alice. In the continuous-time mode theory analysis section, it is known that as long as there exists a signal that can be calibrated to the frequency deviation, a digital signal processing function $f_{\mathrm{dsp}}$ can be obtained so that the matching factor of the quantum signal is 1. Based on the time-varying recovery method, the optimal matching of the two modes of Alice and Bob can be achieved, and the quantum signal can be recovered. Figure~\ref{fig:3-3}(c) shows the CPS of Bob's recovered signal, from which the ASE signal has been recovered to the initial position.

\begin{figure*}[ht]
    \centering
    \includegraphics[width=1\linewidth]{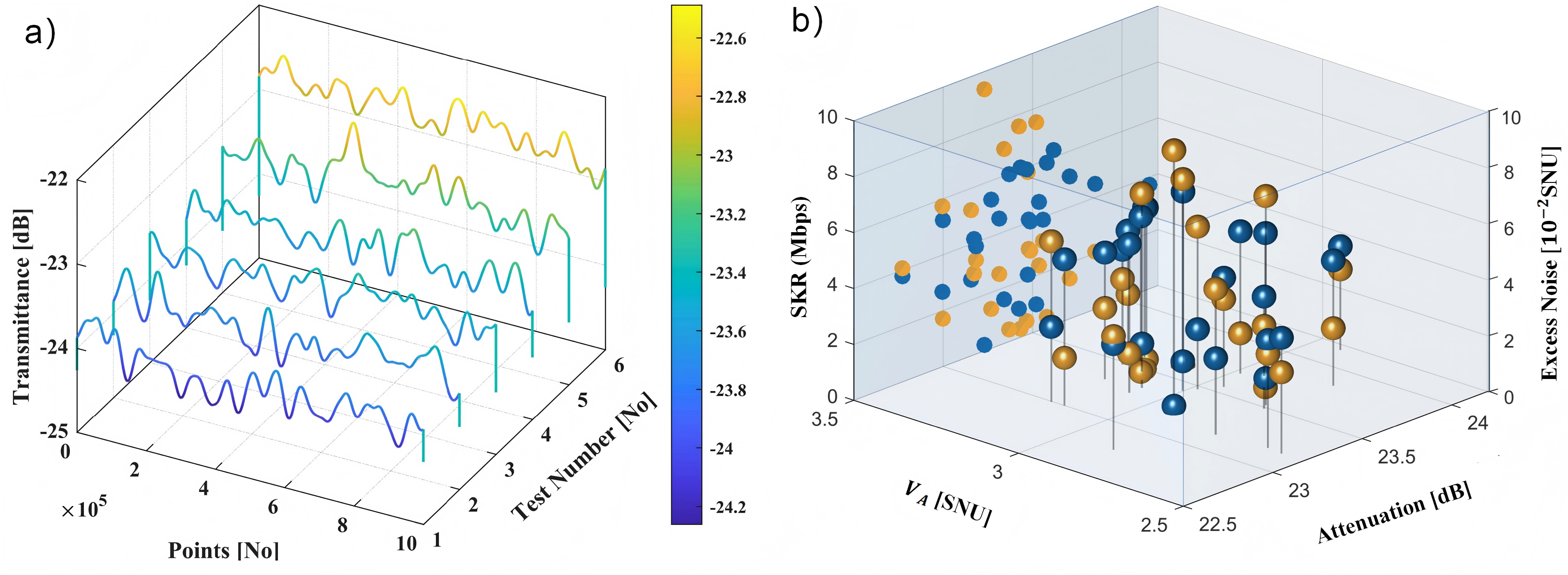}
    \caption{Experimental results in a turbulence-free channel. (a) Time-varying transmittance in a -23.5 dB turbulence-free scenario. (b) Excess noise, modulation variance, transmittance, and secret key rate of 25-frame signals measured in a turbulence-free scenario. The three axes represent the equivalent modulation variance of the signal, the estimated average transmittance, the excess noise, and the secret key rate, respectively. The blue dots indicate excess noise, and the yellow dots indicate secure key rate.}
    \label{fig:3-5}
\end{figure*}

In a free-space channel environment, accurate estimation of channel transmittance is imperative. In our proposed scheme, the beacon light serves a dual function. It can be used not only as a self-referenced pilot to calibrate the thermal signal frequency, but also as a pre-calibrated signal to calibrate the time-varying channel transmittance. The time-varying channel transmittance estimation method described in the theoretical analysis section is employed. In the pre-calibration step, the transceiver and the receiver are connected by an optical fiber of minimal length. Bob detects the signal and recovers it. The recovered beat signal records the information about the thermal signal as it is affected by the physical processes of the transceiver system in the absence of channel attenuation. When a free-space channel is accessed between Alice and Bob, the beacon light traverses the transceiver system via the same physical processes as in the pre-calibration scenario, except that it is additionally affected by the change in channel transmittance. This provides the basis for accurate estimation of time-varying channel transmittance using beacon light. When a free-space channel is accessed between Alice and Bob, the beacon light traverses the transceiver system via the same physical processes as in the pre-calibration scenario, except that it is additionally affected by the change in channel transmittance. This provides the basis for accurate estimation of time-varying channel transmittance using beacon light. Comparing the power of the beat signal at each time point with the pre-calibrated signal allows for accurate estimation of signal transmittance. Figure~\ref{fig:3-5}(a) presents the estimated time-varying transmittance for several frames of the signal in a turbulence-free environment with a channel transmittance of -23.5 dB. Within a single frame, the estimated transmittance is continuously varying, and the variation is less than 1 dB. Compared to a stable channel, the fading channel introduces additional excess noise \cite{fading_noise}, which is given by
\begin{equation}
    \label{eq3.3}
    \varepsilon_{\mathrm{fad}} = Var(\sqrt{\eta})(V_A-1),
\end{equation}
where $\eta$ is the channel transmittance within a single frame and $V_A$ is the equivalent modulation variance of Alice. We calculated the fadding noise of the system using random 30 frames and the mean fadding noise $\overline{\varepsilon_{\mathrm{fad}}} = 0.003$. It should be noted that although each final data frame consists of multiple sampled data frames with transmission variations within 1 dB, the parameter evaluation is performed individually on each sampled data frame. The final parameter evaluation result for the data frame is obtained by taking the average of the parameters from the entire set of sampled data frames. Consequently, the parameter evaluation outcome of the final data frame exclusively reflects the system performance under that particular transmission subchannel. The 1 dB transmission interval does not introduce additional noise.

\begin{table*}
    \centering
    \caption{Key parameters in LLO-PSP-CVQKD scheme.}
    \label{table:1}
    \begin{tabular}{l|cccccccccc}
    \hline
         Parameters & $\eta_0$ & $\eta_{1(3)}$ & $\nu_a^{\mathrm{ele}}$ & $\nu_b^{\mathrm{ele}}$ & $\eta^{a(b)}_{x(p)}$ & $F_m$ & $T_{\mathrm{min}}$ & $T_{\mathrm{max}}$ & $\beta$ & FER\\
    \hline
         Value & 0.0299 & 0.01 & 0.34 & 0.38 & 0.56 & 20 GHz & -23.5 dB & -16 dB & 0.96 & 0.3\\
    \hline
    \end{tabular} 
    \begin{minipage}{\linewidth}
    \raggedright
    ${ }^a$  $\eta_0$ is the transmittance of optical attenuator. $\eta_{1(3)}$ is the transmittance of $\mathrm{BS}_1(\mathrm{BS}_3)$. $\nu_a^{\mathrm{ele}}$ is Alice's (Bob's) electrical noise of the detector. $\eta_{x(p)}^{a(b)}$ is the detection efficiency of Alice’s (Bob’s) heterodyne detector. $F_m$ is the repetition rate. $T_{\mathrm{min}(\mathrm{max})}$ is the minimum (maximum) channel transmittance. $\beta$ is the reconciliation efficiency. FER is the frame error rate of the reconciliation.
    \end{minipage}
    
\end{table*}

\subsection*{Experimental results}
We have realized a high-performance LLO-PSP-CVQKD scheme based on the experimental setup, and the key experimental parameters are summarized in Table~\ref{table:1}. Under the experimental conditions described above, we found that the maximum channel loss the scheme can tolerate is 23.5 dB. The measurement results in an unturbulated free-space channel environment are shown in Figure~\ref{fig:3-5}(b). The average value of the equivalent modulation variance, transmittance excess noise is 2.973 SNU, -23.4, and 0.0393 SNU, respectively.

Subsequently, the key rate in scenarios involving turbulent disturbances will undergo further exploration. The key rate formula is (more details in see Supplementary Information, Note 5):
\begin{equation}
    \label{eq4.1}
    SKR=F_\mathrm{m}(1-\mathrm{FER})\left[\beta I_{\mathrm{AB}}-\kappa_{\mathrm{BE}}\right],
\end{equation}
\noindent where $F_m$ is the repetition frequency of the system, FER is the frame error rate of reverse reconciliation, $\beta$ is the reconciliation efficiency, $I_{\mathrm{AB}}$ is the Shannon mutual information between Alice and Bob, $\kappa_{\mathrm{BE}}$ is the Holevo bound of the information between Bob and Eve. In the proposed PSP-CVQKD scheme, the ICR detector operates within a detection bandwidth of 20 GHz, while the oscilloscope sampling frequency is 40 Gs/s. The repetition frequency of the system is determined by the minima of both the detector bandwidth and the sampling frequency, resulting in $F_m = 20$ GHz. According to Eq.~\ref{eq4.1}, the average SKR in the 23.5 dB loss channel is 3.3492 Mbps. This characteristic of the PSP-CVQKD protocol distinguishes it from the ASP-GMCS-CVQKD scheme. In the case of transmittance jitter, the total excess noise of the system can be calculated by introducing the probability of transmittance $P_T$, which is described by the equation:
\begin{equation}
    \label{eq4.2}
    R_{\mathrm{total}}=\sum_{T} P_T\cdot R_T,
\end{equation}
\noindent where $R_T$ is the SKR at different transmittance calculated by the Eq.~\ref{eq4.1}.

\begin{figure*}[ht]
    \centering
    \includegraphics[width=\linewidth]{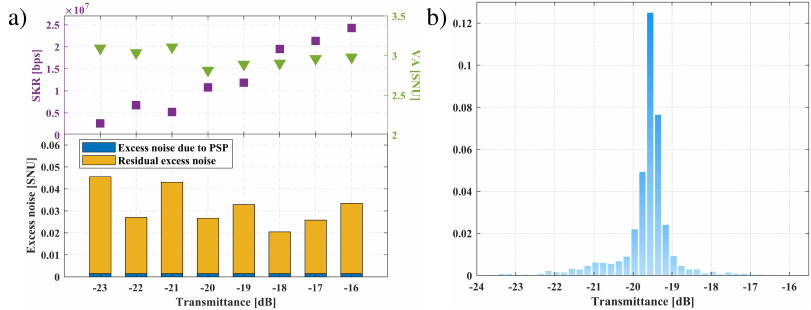}
    \caption{Experimental results under turbulent channels. (a) Mean excess noise and its composition, equivalent modulation variance, and secret key rate at different transmittances. (b) Probability distribution of channel transmittance.}
    \label{fig:4-1}
\end{figure*}

We statistically analyze the average transmittance of 2000 signal frames recorded within 100 ms. Turbulent interference causes the signal transmittance to vary from -25 to -16 dB. The probability distribution of transmittance is shown in ~Figure~\ref{fig:4-1}(a). According to the a priori experimental results, our system can effectively operate with channel attenuation greater than 24 dB. Therefore, we choose signals with transmittance ranging from -16 dB to -24 dB for further analysis.

Signals are grouped at a 1 dB transmittance interval. As illustrated in ~Figure~\ref{fig:4-1}(b), the mean excess noise, the equivalent modulation variance, and the secret key rate of the signals corresponding to each group are displayed in three panels from top to bottom. The x-axis is defined as the initial value of each transmittance interval. The equivalent modulation variance of the signal fluctuates slightly around 3 SNU. In addition, we analyzed the composition of the average excess noise for each transmittance condition. The results indicate that the PSP noise is not a major component in this scheme, with an average value of merely 0.0014 SNU. This phenomenon is attributed to implementing a 99:1 BS instead of a 50:50 BS at Alice, which is equivalent to increasing the attenuation at Alice, and thus effectively suppresses the PSP noise. Within the (-17 dB, -16 dB] range, the average key rate can approach 24.305 Mbps. The mean key rate within the (-24 dB, -23 dB] range is 2.596 Mbps. Considering the probability of occurrence of the transmittance, the average secret key rate calculated according to the Eq.~\ref{eq4.2} is 10.342 Mbps in the current case with turbulent disturbances.

\section*{DISCUSSION}

In this paper, we propose and experimentally realize a free-space LLO-CVQKD scheme with thermal source. We introduce beacon optical calibration of the center frequency of the thermal state signal, which overcomes the problem that the thermal state itself is unable to complete the LO. First, we perform a theoretical derivation based on the temporal modes of continuous-mode states, and then we experimentally verify this LLO-PSP-CVQKD system. The placement of the LO at the receiver greatly simplifies the experimental optical path compared to the TLO scheme, which is more suitable for applications in remote scenarios. Finally, using the self-referenced pilot scheme, the system can achieve lower excess noise with time-varying frequency offsets and phase drifts through time-varying parameter estimation, and can realize time-varying estimation of channel transmittance in a much simpler way, which was not possible with the previous free-space PSP-CVQKD scheme. The final SKR achieves 10.342 Mbps over a turbulent free-space channel with a maximum loss of 23.5 dB. This work sets the foundation for future applications regarding PSP schemes in long-range scenarios.

It should be mentioned here that the experiment is a proof-of-principle demonstration for the long-distance PSP scheme. Due to the fluctuation of transmittance and the limit of the acquisition equipment, collecting a large number of symbols remains crucial, and only the asymptotic is considered. This can be achieved by employing a custom data acquisition card (DAQ) with more memory instead of the oscilloscope to collect data. Moreover, offline data processing greatly increases the time required for parameter estimation, which may result in an inability to track system changes on time. Real-time implementation can be achieved by using high-performance PCBs for post-processing or graphics processing unit \cite{SA100km}. On the other hand, the high attenuation of the free-space channel is most caused by the low coupling efficiency, which is a much simplifier condition than the actual field trial where channel attenuation is caused by various atmospheric effects. For field implementation, atmospheric effects compensation should be considered in the system.

In summary, this experiment demonstrated the feasibility of the LLO-PSP-CVQKD scheme, and it has the potential of a high-rate QKD system over free space. Given the advantages of the PSP-CVQKD system in terms of ease of integration and low power consumption, we believe that, if the light source could be further optimized, such as by using sunlight, the system could serve as a key component of an inter-satellite QKD network.


\section*{MATERIALS AND METHODS}

\subsection*{\label{Methods: 1} Analysis of excess noise due to passive state preparation}

The LLO-PSP-CVQKD excess noise model can be similarly derived with reference to the transmitted LO PSP-CVQKD model. To simplify the derivation process, only X is analyzed below, and P is analyzed similarly to X. As shown in Figure~\ref{fig:sche}, after the optical attenuator, the thermal signal transmitted into the channel is
\begin{equation}
    \label{eqApp3.1}
    \begin{split}
        x_1 =& \sqrt{\eta_0\eta_1}x_{\mathrm{in}}-\sqrt{\eta_0(1-\eta_1)}x_{\nu_1}\\
        &-\sqrt{\eta_0\eta_3}x_{\nu_3}-\sqrt{1-\eta_0}x_{\nu_4},
    \end{split}
\end{equation}
where $x_{\mathrm{in}}$ is the X-quadrature of the thermal source output mode. $\eta_0$, $\eta_1$, and $\eta_3$ are the transmittances of the optical attenuator $\mathrm{BS}_3$, $\mathrm{BS}_1$ and $\mathrm{BS}_3$, respectively. $x_{\nu_1}$, $x_{\nu_3}$, and $x_{\nu_4}$ are the vacuum noise introduced by $\mathrm{BS}_1$, $\mathrm{BS}_3$, and the attenuator, respectively. The vacuum noises all satisfy $\left\langle x_\nu^2\right\rangle = 1$. The X-quadrature of the Alice detected is
\begin{equation}
    \label{eqApp3.2}
    \begin{split}
        x_2 =& \sqrt{\frac{\eta_x^a(1-\eta_1)}{2}}x_{\mathrm{in}} - \sqrt{\frac{\eta_x^a\eta_1}{2}}x_{\nu_1} + \sqrt{\frac{\eta_x^a}{2}}x_{\nu_2}\\
        &+\sqrt{\frac{\eta_x^a(1-\eta_3)}{2}}x_{\nu_3}-\sqrt{1-\eta_x^a}x_{\nu_a} + E_a,
    \end{split}
\end{equation}
where $\eta_x^a$ is the detection efficiency of the heterodyne detector, and $\eta_x^a=\eta_p^a$. $x_{\nu_2}$ is the vacuum noise introduced by the frequency mixer. $x_{\nu_a}$ and $E_a$ are vacuum noise and electrical noise caused by the detector, respectively. We define the electrical noise variance as $\nu^{\mathrm{ele}}_a=\left\langle {E_a}^2\right\rangle$. Given $x_2$, Alice's optimal estimate of $x_1$ is $x_{\mathrm{opt}}=\alpha_{1,2}x_2$, where $\alpha_{1,2}=\frac{\left\langle x_1x_2\right\rangle}{\left\langle x_2^2\right\rangle}$. Thus, we can determine $\alpha_{1,2}$ is 
\begin{equation}
    \label{eqApp3.3}
    \begin{split}
        \alpha_{1,2}
        =\frac{2(n_0+1)\sqrt{2\eta_0\eta_1\eta_x^a(1-\eta_1)}-\sqrt{2\eta_0\eta_3\eta_x^a(1-\eta_3)}}{n_0\eta_x^a(1-\eta_1)+\eta_x^a(1-\eta_3)+2\nu^{\mathrm{ele}}_a+2},
    \end{split}
\end{equation}
where $n_0$ is the average photon number of the thermal source, and satisfies the relation $\left\langle x_{\mathrm{in}}^2\right\rangle=2n_0+1$. The excess noise of state preparation is defined $\varepsilon_{\mathrm{psp}} =\left\langle (x_1-\alpha_{1,2}x_2)^2\right\rangle$. Therefore, we can obtain
\begin{equation}
    \label{eqApp.3.4}
    \begin{split}
        \varepsilon_{\mathrm{psp}} = 
        &\frac{-\eta_x^a\left[ 2(V_A+\eta_0)\sqrt{\eta_1(1-\eta_1)}-\eta_0\sqrt{\eta_3(1-\eta_3)}\right]^2}{2V_A\eta_x^a(1-\eta_1)+\eta_0\eta_x^a(1-\eta_3)+2\eta_0(\nu^{\mathrm{ele}}_a+1)}\\
        &+2V_A\eta_1+\eta_0\eta_3.\\
    \end{split}
\end{equation}

\subsection*{Time-varying compensation and transmittance estimation}

Based on the continuous-time mode theory for LLO-PSP-CVQKD scheme, after Bob performed coherent detection on the received signal, the beat signal is an easily distinguishable and prominent shock function in the spectrum. This beat signal conveys information about the frequency bias and phase noise, and the signal can be compensated for using the self-referenced LO transmitted from Alice. First, let the DSP function work like a filter $f_{\mathrm{dsp1}}^q=H_p^m$, where $H_p^m$ is the impulse response function of the bandpass filter on the $m$-th data. Thus, we can separate the beat signal $\left| \sqrt{\eta_0T(t)}\gamma_\mathrm{L1}\right\rangle_{\xi_\mathrm{L1}^c}$ from the received signal, which matches the band of the bandpass filter $H_p^m$. Thus, the mode-matching factor $\sqrt{\eta_q^1}$ of the beat signal is
\begin{equation}
    \label{eq2.15}
    \begin{split}
        \sqrt{\eta_q^1}&=\int_{t_0}^{t_0+N/F_m}dt\frac{1}{\sigma_{\mathrm{cal}}^*}\xi_{\mathrm{LO}}^*(t)G_{\mathrm{dsp1}}^{N,q^*}\xi_\mathrm{L1}^c(t)\\
        &=\int_{t_0}^{t_0+N/F_m}dt\frac{1}{\sigma_{\mathrm{cal}}^*}G_{\mathrm{dsp1}}^{N,q^*}\xi_{\mathrm{LO}}^{0^*}(t)\xi_\mathrm{L1}^0(t)e^{\mathrm{j}\left(\omega_{\mathrm{LO}}^tt-\omega_st-\theta_{\mathrm{LO}}^t+\theta_c^t\right)}.
    \end{split}
\end{equation}
The first-order moment $\mu_q^1$ of the beat signal is:
\begin{equation}
    \label{eq2.16}
    \begin{split}
        \mu_q^1=&\left\langle \sqrt{\eta_0T(t)}\gamma_\mathrm{L1}\right|\hat{I}_{\Xi_{\mathrm{dsp}}^q}^{\mathrm{SNU}}\left| \sqrt{\eta_0T(t)}\gamma_{\mathrm{L1}}\right\rangle_{\xi_\mathrm{L1}^c}\\
        =& \sqrt{\eta_e\eta_0T(t)}\gamma_{\mathrm{L1}} e^{\mathrm{j}\left(\omega_{\mathrm{LO}}^tt-\omega_st-\theta_{\mathrm{LO}}^t+\theta_c^t\right)},
    \end{split}
\end{equation}
which carries all the information about the time-varying frequency offset and the phase noise, as well as the channel phase noise. The next DSP step $f_{\mathrm{dsp2}}^q$ is the operation of amplitude normalization of $\mu^1_q$ to obtain the new first-order moment $\mu_q^2$ as:
\begin{equation}
    \label{eq2.17}
    \mu_q^2=e^{\mathrm{j}\left(\omega_{\mathrm{LO}}^tt-\omega_st-\theta_{\mathrm{LO}}^t+\theta_c^t\right)}.
\end{equation}

Now we can get the DSP algorithm $f_{\mathrm{dsp3}}^q=\mu_q^2$. This indicates that the signal undergoes the correct rotation so that the time-varying frequency offset and phase noise are compensated, and the new mode-matching coefficient and first-order moment are:
\begin{equation}
\label{eq2.18}
    \begin{split}
        \sqrt{\eta_q^3}=&\int_{t_0}^{t_0+N/F_m}dt\frac{1}{\sigma_{\mathrm{cal}}^*}\xi_{\mathrm{LO}}^*(t)G_{\mathrm{dsp3}}^{N,q^*}\xi_\mathrm{L1}^c(t)\\
        =&\int_{t_0}^{t_0+N/F_m}dt\frac{1}{\sigma_{\mathrm{cal}}^*}\xi_{\mathrm{LO}}^{0^*}(t)\xi_\mathrm{L1}^0(t),
    \end{split}
\end{equation}

\begin{equation}
\label{eq2.19}
    \begin{split}
        \mu_q^3
        =& \sqrt{\eta_e\eta_0T(t)}\gamma_\mathrm{L1}.
    \end{split}
\end{equation}
Therefore, the DSP operation to realize the optimal reception of the quantum signal is:
\begin{equation}
\label{eq2.20}
    \begin{split}
        f_{\mathrm{dsp}}^{N,\mathrm{opt}}=\frac{\xi_i^{0^*}(t)}{\xi_{\mathrm{LO}}^{0^*}(t)}\mu_q^2\overset{CW}{=}\mu_q^2.
    \end{split}
\end{equation}
Ultimately, the mode-matching factor and first-order moment of the quantum signal detected by Bob are
\begin{equation}
\label{eq2.21}
    \begin{aligned}
        \sqrt{\eta_B}=&\int_{t_0}^{t_0+N/F_m}dt\frac{1}{\sigma_{\mathrm{cal}}^*}\xi_{\mathrm{LO}}^*(t)G_{\mathrm{dsp}}^{N,\mathrm{opt}^*}\xi_i^c(t)\\
        =&\int_{t_0}^{t_0+N/F_m}dt\frac{1}{\sigma_{\mathrm{cal}}^*}\xi_{\mathrm{LO}}^{0^*}(t)\xi_i^0(t)\frac{{\xi_i^{0^*}}(t)}{\xi_{\mathrm{LO}}^{0^*}(t)}=1,
    \end{aligned}
\end{equation}

\begin{equation}
\label{eq2.22}
    \begin{split}
        \mu_B
        =& \left\langle \sqrt{\eta_0T(t)}\left(x_b+jp_b\right)\right|\hat{I}_{\Xi_{\mathrm{dsp}}^B}^{\mathrm{SNU}}\left| \sqrt{\eta_0T(t)}\left(x_b+jp_b\right)\right\rangle_{\xi_i^c}\\
        =& \sqrt{\eta_e\eta_0T(t)}\left(x_b+jp_b\right)\\
        =& \sqrt{K\eta_e\eta_0T(t)}\left(x_a+jp_a\right),
    \end{split}
\end{equation}

\noindent where $K=\frac{1-T_1}{T_1}$. After time-varying frequency offset and phase recovery, $\mu_B=\sqrt{K\eta_0T(t)}\cdot \mu_A$, the recovered signal at Bob is a constant multiple of the scaling of the signal detected at Alice. It can be regarded that the received signal detected by Bob and the transmitted signal detected by Alice are matched.
In addition, the beat signal is processed with the same DSP ($f_{\mathrm{dsp1}}^q$, $f_{\mathrm{dsp2}}^q$, $f_{\mathrm{dsp3}}^q$) in the pre-calibration phase to obtain $\mu_{q,0}^3$:
\begin{equation}
\label{eq2.23}
    \mu_{q,0}^3=\sqrt{\eta_e\eta_0}\gamma_\mathrm{L1}.
\end{equation}
The time-varying transmittance can be obtained as follows:
\begin{equation}
\label{eq2.24}
    T(t)=\frac{|\mu_q^3|^2}{|\mu_{q,0}^3|^2}=\frac{\eta_e\eta_0T(t)\gamma_\mathrm{L1}^2}{\eta_e\eta_0\gamma_\mathrm{L1}^2}.
\end{equation}

In summary, we demonstrate the feasibility of the proposed LLO-PSP-CVQKD scheme in continuous-time mode theory and prove the corresponding time-varying compensation and time-varying transmittance estimation methods.




\clearpage 

%
\bibliography{main} 
\bibliographystyle{sciencemag}

%
%
%
%
%
%


\section*{Acknowledgments}
We thank our colleagues for their contributions to the work cited.
\paragraph*{Funding:}
This work was supported by Quantum Science and Technology-National Science and Technology Major Project (Grant No. 2021ZD0300703), Shanghai Municipal Science and Technology Major Project (2019SHZDZX01), and the Key R\&D Program of Guangdong province (Grant No. 2020B0303040002), and the National Natural Science Foundation of China (No. 62101320), Open Research Fund of the State Key Laboratory of Photonics and Communications (Grant No. 2026QZKF023).
\paragraph*{Author contributions:}
G.Z. conceived the research project. H.Y. and P.H. designed the scheme with assistance from Y.X. and T.W.. H.Y. and X.L. carried out the experiments. X.L. contributed the approach to the data processing. Y.X. contributed the approach to post-processing. H.Y., X.L. and P.H. wrote the manuscript with contributions from all authors.
\paragraph*{Competing interests:}
There are no competing interests to declare.
\paragraph*{Data and materials availability:}
All data needed to evaluate the conclusions in the paper are present in the paper and/or the materials cited herein. Additional data related to this paper may be requested from the authors.

\end{document}